\documentclass[journal, final]{IEEEtran}
\usepackage{xcolor}
\usepackage{cite}
\usepackage{textcomp}
\usepackage{lipsum}
\usepackage{mwe}
\usepackage{rotating}
\usepackage{lineno}
\usepackage{amssymb}
\usepackage{amsmath}
\usepackage{amsthm}
\usepackage{epstopdf}
\usepackage{graphicx}
\usepackage{multirow}
\usepackage{tabls}
\usepackage{ctable}
\usepackage{algorithm}
\usepackage{algpseudocode}
\usepackage[utf8]{inputenc}
\usepackage[hidelinks]{hyperref}
\usepackage[english]{babel}
\DeclareGraphicsExtensions{.eps}
\usepackage{subfigure}
\usepackage{rotating}
\usepackage{lineno}
\usepackage{amssymb}
\usepackage{amsmath}
\usepackage{amsthm}
\usepackage{epstopdf}
\usepackage{graphicx}
\usepackage{caption}
\usepackage{multirow}
\usepackage{tabls}
\usepackage{ctable}
\usepackage{algorithm}
\usepackage{algpseudocode}
\usepackage[utf8]{inputenc}
\usepackage[hidelinks]{hyperref}
\usepackage[english]{babel}

\newcommand{\sigmah}{\sigma_h}
\newcommand{\hjeff}{\left | h_{j,\text{eff}}  \right |^2}
\newcommand{\ehjeff}{h_{j,\text{eff}}}
\newcommand{\hkeff}{\left |h_{k,\text{eff}}  \right |^2}
\newcommand{\ehkeff}{h_{k,\text{eff}}}
\newcommand{\egmeff}{g^m_{j,\text{eff}}}
\newcommand{\egoeff}{g^1_{j,\text{eff}}}

\newcommand{\egMeff}{g^M_{j,\text{eff}}}

\newcommand{\hzeroeff}{\left |h_{j,\text{eff}}  \right |}

\newcommand{\gone}{g_1}
\newcommand{\gm}{g_m}
\newcommand{\gM}{g_M}
\newcommand{\goneeff}{\left |g_{ 1,\text{eff}}  \right |}

\newcommand{\gtwoeff}{\left |g_{ 2,\text{eff}}  \right |}

\newcommand{\gMeff}{\left |g_{ M,\text{eff}}  \right |}
\newcommand{\muI}{\mu_I}
\newcommand{\sigmaI}{\sigma_I}
\newcommand{\pnoma}{P}
\newcommand{\petotalnoma}{P_{\scriptscriptstyle e,overall}^{\scriptscriptstyle (NOMA)}}
\newcommand{\petotal}{P_{\scriptscriptstyle e,overall}}
\newcommand{\pegssk}{P_{\scriptscriptstyle e,GSSK}}

\newcommand{\penoma}{P_{\scriptscriptstyle e,C-NOMA}^{\scriptscriptstyle (m)}}

\newcommand{\nt}{n_{t}}
\newcommand{\Nt}{N_{t}}
\newcommand{\rategssk}{R_{\scriptscriptstyle GSSK}}
\newcommand{\ratenoma}{R_{\scriptscriptstyle NOMA}}
\newcommand{\ratengssk}{R_{\scriptscriptstyle N-GSSK}}

\newcommand{\mh}{M_{H}}
\newcommand{\rhodash}{\rho'}
\newcommand{\varn}{\sigma_n^2}
\newcommand{\varnnoma}{\sigma_n}

\newcommand{\deltaone}{\delta_1}
\newcommand{\deltam}{\delta_m}
\newcommand{\deltak}{\delta_k}
\newcommand{\etam}{\eta_m}
\newcommand{\uezero}{\text{UE}_{\scriptscriptstyle 0}}
\newcommand{\ueone}{\text{UE}_{\scriptscriptstyle 1}}
\newcommand{\uetwo}{\text{UE}_{\scriptscriptstyle 2}}
\newcommand{\ueg}{\text{UE}_{\scriptscriptstyle M}}
\newcommand{\uegone}{\text{UE}_{\scriptscriptstyle 1}}
\newcommand{\uegtwo}{\text{UE}_{\scriptscriptstyle 2}}
\newcommand{\uegmminusone}{\text{UE}_{\scriptscriptstyle m-1}}
\newcommand{\uegm}{\text{UE}_{\scriptscriptstyle M}}
\newcommand{\uegmplusone}{\text{UE}_{\scriptscriptstyle m+1}}
\newcommand{\uegmplustwo}{\text{UE}_{\scriptscriptstyle m+2}}

\newcommand{\yE}{y_{\scriptscriptstyle E}^{\scriptscriptstyle (0)}}
\newcommand{\erf}{\text{erf}}

\newtheorem{Proposition}{Proposition}

\title{Interplay Between NOMA and GSSK: Detection Strategies and Performance Analysis}
\author{Sanjeev~Gurugopinath,~\IEEEmembership{Member,~IEEE},~Sami~Muhaidat,~\IEEEmembership{Senior~Member,~IEEE},\\Rajaleksmi~Kishore,~\IEEEmembership{Student~Member,~IEEE},~Paschalis~C.~Sofotasios,~\IEEEmembership{Senior~Member,~IEEE},\\Faissal~El~Bouanani,~\IEEEmembership{Senior~Member,~IEEE},~and~Halim~Yanikomeroglu,~\IEEEmembership{Fellow,~IEEE} 
\thanks{S.~Gurugopinath is with the Department of Electronics and Communication Engineering, PES University, Bengaluru 560085, India, (email: {\rm sanjeevg@pes.edu}).}
\thanks{S.~Muhaidat is with the KU Center for Cyber-Physical Systems, Department of Electrical and Computer Engineering, Khalifa University, Abu Dhabi 127788, UAE, (email: {\rm muhaidat@ieee.org}).}
\thanks{R.~Kishore is with the Department of Electrical and Electronics Engineering, BITS Pilani, K.~K.~Birla Goa Campus, Goa 403726, India, (email: {\rm lekshminair2k@yahoo.com}).}
\thanks{P.~C.~Sofotasios is with the Center for Cyber-Physical Systems, Department of Electrical and Computer Engineering, Khalifa University, Abu Dhabi 127788, UAE, and also with the Department of Electrical Engineering, Tampere University, Tampere 33101, Finland,  (email: {\rm p.sofotasios@ieee.org}).}
\thanks{F.~El Bouanani is with ENSIAS College of Engineering, Mohammed V University, Rabat, Morocco, (email: {\rm f.elbouanani@um5s.net.ma}).}
\thanks{H.~Yanikomeroglu is with the Department of Systems and Computer Engineering, Carleton University, Ottawa, ON K1S 5B6, Canada (e-mail: {\rm halim@sce.carleton.ca}).}
}
 
\begin{document}
\captionsetup[figure]{name={Fig.},labelsep=period}
 \maketitle

\begin{abstract}
Non-orthogonal multiple access (NOMA) is a technology enabler for the fifth generation and beyond networks, which has shown a great flexibility such that it can be readily integrated with other wireless technologies.  In this paper, we investigate the interplay between NOMA and generalized space shift keying (GSSK) in a hybrid NOMA-GSSK (N-GSSK) network.  Specifically, we provide a comprehensive analytical framework and  propose a novel suboptimal energy-based maximum likelihood (ML)  detector  for the N-GSSK scheme. The proposed ML decoder exploits the energy of the received signals in order to estimate the active antenna indices. Its performance is  investigated in terms of pairwise error probability, bit error rate union bound, and achievable rate. Finally, we establish the validity of our analysis through Monte-Carlo simulations and demonstrate that N-GSSK outperforms conventional NOMA and GSSK, particularly in terms of spectral efficiency.
\end{abstract}


\IEEEpeerreviewmaketitle

\section{Introduction} \label{sec:introduction}
The unprecedented growth of mobile data traffic and the massive number of connected devices, due to the emergence of internet of things (IoT), have posed several challenges for the fifth generation and beyond networks, such as high spectral efficiency, massive connectivity, and requirements for low latency. Accordingly, several promising technologies have been proposed to address these stringent requirements, including massive multiple-input multiple-output (MIMO) \cite{Liu_IEEEJSAC_2017, Silva_TCOM_2020}, millimetre wave (mmWave) communications \cite{Shao_TCOM_2020,Rappaport_IEEETAP_2017,Rappaport_IEEEACe_2013,Al-Ogaili_IEEEconf_2016}, and spatial modulation (SM) techniques \cite{Hendraningrat_Access_2020}, which enable information transmission through spatial and signal constellation \cite{Mesleh_IEEEVT_2008,jaydeepan_IEEETWC_2009,Renzo_IEEEVT_2011}. 

As a key enabling technology for the fifth generation and beyond networks networks, non-orthogonal multiple access (NOMA) is envisioned to increase the system throughput and support massive connectivity \cite{Silva_TCOM_2020}. It is worth noting that NOMA can be classified into two different approaches: (a) power domain \cite{Islam_IEEECST_2017}, and (b) code domain \cite{Le_IET_2018}. In the power domain NOMA, users are assigned different power levels over the same time and frequency resources. On the contrary, in the code domain NOMA, multiplexing is carried out using spreading sequences with low cross-correlation, similar to code division multiple access technology.\footnote{In this work, by NOMA, we refer to the power domain NOMA.}
The basic principle of NOMA is to allow multiple users to share the same frequency and time resources while  controlling the level of inter-user interference \cite{Islam_IEEECST_2017}. Unlike conventional orthogonal multiple access (OMA) techniques, where  users in a cell are assigned dedicated communication resources, e.g., time, frequency or code, NOMA employs superposition coding, where multiple users are multiplexed in the power domain at the transmitter side. At the users’ terminals, multi-user detection is realized by successive interference cancellation (SIC).  It has been demonstrated in the recent literature that NOMA outperforms OMA in several aspects such as spectral efficiency, which is realized by serving multiple users at the same time and frequency resource block, interference mitigation through SIC, and support for massive connectivity \cite{Wan_IEEEWC_2018}. Furthermore, users in NOMA do not require a prescheduled time slot structure; hence, NOMA offers lower latency. Moreover, NOMA can ensure user-fairness and diverse quality-of-service by employing flexible power control strategies between  \emph{strong} and \emph{weak} users \cite{Islam_IEEECST_2017}.

On the other hand, SM, which is considered as a promising MIMO technique for the fifth generation and beyond networks wireless networks, has received significant attention in the recent literature. In SM, information bits are divided into blocks, each of which consists of two subblocks. The first subblock maps the information bits to an arbitrary signal constellation, whereas the second subblock determines the index of the transmit antenna to be used for transmission \cite{Jeganathan_IEEECL_2008}.   
In order to reduce the system’s complexity, space shift keying (SSK) modulation, which is a variant of spatial modulation, was introduced in \cite{Jeganathan_IEEETWC_2009}.  The key idea of SSK is the activation of one antenna at each symbol duration, which cleverly sends source information to a receiver while removing the effects of inter-antenna interference. Since SSK avoids modulation/demodulation of data symbols, the complexity of the system is significantly reduced but at the cost of a reduced spectral efficiency. To address this concern, generalized space shift keying (GSSK) modulation was proposed in \cite{Jeganathan_PIMRC_2008}, \cite{Su_IEEECL_2015}. Unlike SSK, GSSK allows for more than one active antenna at each symbol duration, resulting in a better spectral efficiency but at the cost of additional receiver complexity.

The integration of NOMA with other physical layer techniques  such as cooperative communication \cite{Kim_IEEECL_2015,Ding_IEEECL_2015,Kara_IEEECL_2019,Ding_IEEEWCL_2016,Liu_IEEESAC_2016} and MIMO\cite{Sun_IEEECL_2015}, has received a significant attention recently. In \cite{Liu_IEEESAC_2016}, a new simultaneous wireless information and
power transfer NOMA scheme was proposed. NOMA-enabled cognitive radio networks and NOMA in vehicle-to-vehicle massive MIMO channels were investigated in \cite{Zhou_IEEEWC_2018} and \cite{Chen_IEEEJSAC_2017}, respectively. The aforementioned studies have focused on the outage and sum rate analysis, and demonstrated that NOMA outperforms other conventional OMA schemes. 

\begin{figure*}[t]
	\centering
	\includegraphics[width=4in, height=2in]{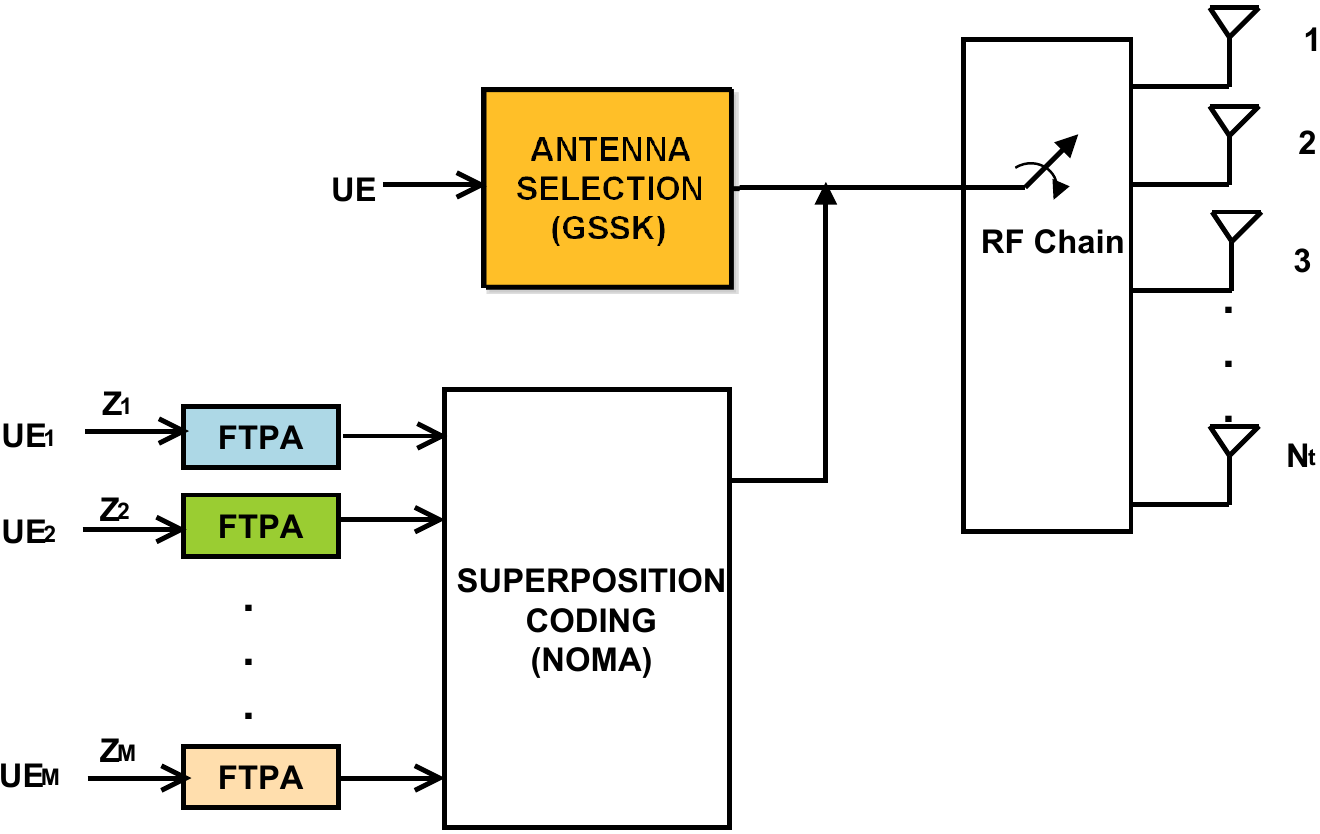} 
	\caption{Transmitter model of N-GSSK.} \label{NOMATX}
\end{figure*}

\subsection{Related Work } \label{LitSurvey}
Extensive research efforts have been made towards investigating the performance of NOMA from different perspectives, and under various scenarios. Liu \textit{et al}.~\cite{Liu_IEEETWC_2018_1} studied the performance of a heterogeneous network with coordinated joint transmission-NOMA in order to enhance the performance of the farthest user in a cell. This scheme was shown to significantly enhance the coverage and throughput performances of all users, especially in dense networks. Ali \textit{et al.}~\cite{Ali_IEEEAccess_2016} formulated a joint  optimization  problem  for the sum-throughput  maximization under several constraints, i.e.,  transmission power budget, minimum rate  requirements,  and  operational SIC requirements. The work in \cite{Liu_IEEETWC_2018} focused on  throughput enhancement. The authors in \cite{Wu_IEEETVT_2018} demonstrated that NOMA requires a longer downlink wireless energy transfer time duration than TDMA, which implies that NOMA is less energy efficient, particularly in those scenarios where energy consumption is of great importance \cite{Hedayati_IEEEVT_2018}. The authors in \cite{Song_IEEEaccess_2018} investigated a system which incorporates both NOMA and OMA in a unified framework. It was shown that  the proposed NOMA-OMA scheme outperforms  NOMA- and OMA-only in terms of spectral and energy efficiency trade-off and user fairness. However,  it suffers from  a level of inter-user interference and relatively high computational complexity. 
\begin{figure*}[t]
	\centering
	\includegraphics[width=6.5in, height=3in]{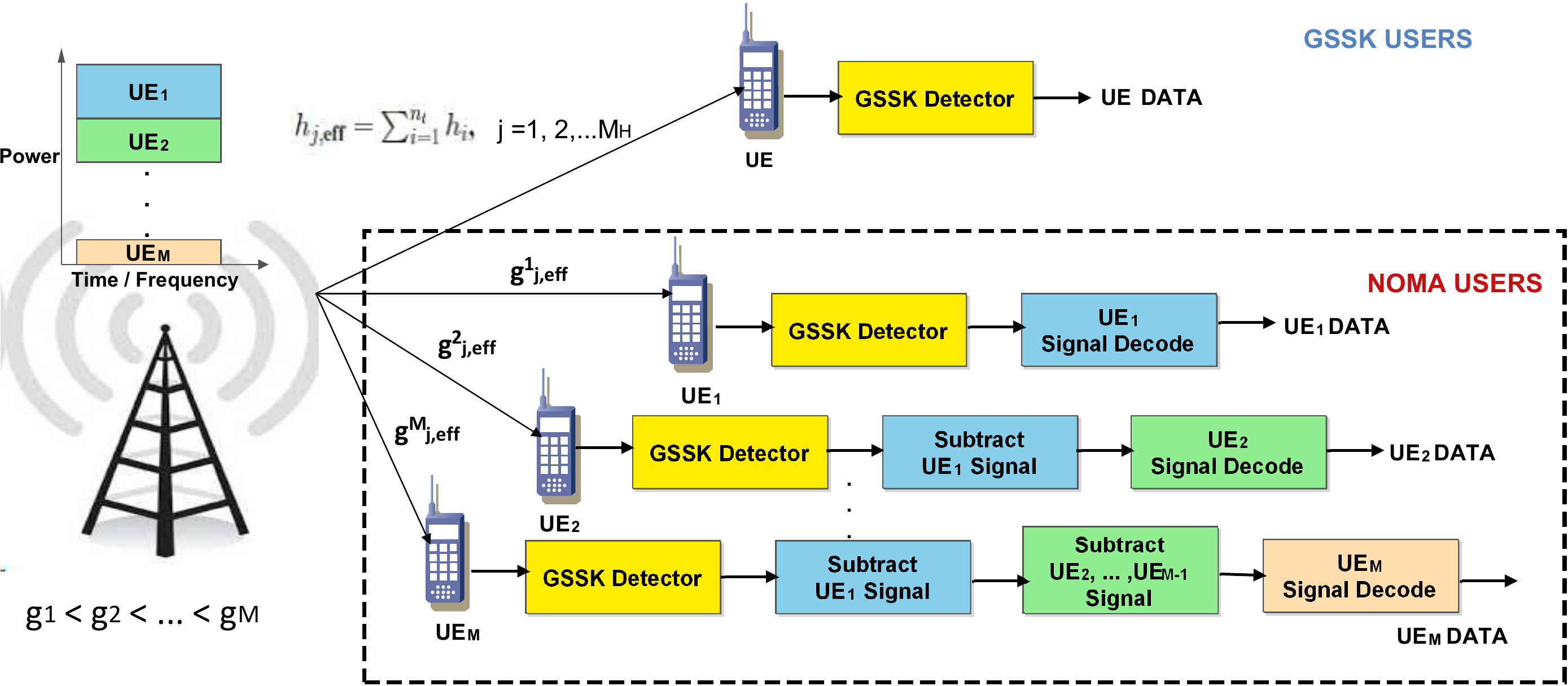} 
	\caption{Receiver model of N-GSSK.} \label{NOMARX}
\end{figure*}

The integration of NOMA with SM has  been recently investigated in the literature \cite{Hendraningrat_Access_2020}. In \cite{Siregar_IEEEConf_2017}, a hybrid detection technique was introduced, which combines NOMA and SM in an uplink transmission scenario. It was shown that the combination of SM and NOMA offers promising spectral efficiency enhancements. Likewise, spectral efficiency analysis was  considered in \cite{Zhong_IEEEWC_2018,Li_IEEEJSTSP_2019} in order to characterize the performance of spatially-modulated NOMA   under different transmission scenarios. More recently, the work of  \cite{Zhu_IEEEaccess_2017} proposed a NOMA-SM system in order to quantify the  trade-off between spectral efficiency and interference mitigation. User pairing for NOMA-SM was also addressed in \cite{Zhu_IEEEaccess_2017}, where the performance of NOMA-SM was compared with OMA-SM as well as transmit antenna grouping-based SM (TAG-SM). The performance gains of NOMA-SM  was further quantified. However, in the aformentioned works, the number of users in a resource block was restricted to two.

The combination of NOMA with SSK was considered in \cite{Irfan_Conf_2015} for a three-user scenario, where a cell edge user was served using SSK modulation and  the other two users served using NOMA. This work was generalized in \cite{Kim_IEEE_CL_2018} by integrating NOMA with GSSK to accommodate more users. It was shown that  NOMA-GSSK  achieves  higher spectral efficiency and lower bit error rate (BER) when compared with conventional GSSK (c-GSSK), conventional NOMA (c-NOMA), and NOMA-SSK systems. This is primarily due to the fact that  users are multiplexed in  power and spatial domain.

\subsection{Motivation}
To the best of authors’ knowledge, the only existing work that combines NOMA with GSSK systems was reported in \cite{Kim_IEEE_CL_2018}, which is called here ideal-NOMA-GSSK (iN-GSSK). Although this work is interesting and paves the way for further advancements in this area, it has the following limitations. 
First, the maximum likelihood (ML) detector at the cell-edge users (GSSK users) is designed under the assumption of perfect knowledge of NOMA signals. Consequently, the introduced detector and the corresponding analysis turned out to be similar to that of  c-GSSK. Second, the proposed framework assumes  perfect SIC. It also assumes   perfect knowledge of  active antenna indices at NOMA users. These are clearly unrealistic assumptions, particularly in practical scenarios.
Third, the detrimental effect of fading on the performance of both GSSK and NOMA users was not discussed. 

In summary, the work in  \cite{Kim_IEEE_CL_2018} treats NOMA and GSSK users independently and does not provide a holistic view of the practical considerations on the integration of NOMA and GSSK schemes.

\subsection{Contributions}
Motivated by the limitations of iN-GSSK \cite{Kim_IEEE_CL_2018} and the  lack of a general theoretic framework for NOMA-GSSK (henceforth called  N-GSSK), we introduce, in this work, a framework for the design and analysis of N-GSSK over fading channels. In particular, we introduce  an analytical framework that investigates the performance of N-GSSK by explicitly multiplexing information in the spatial and power domains. We relax the assumption of  perfect knowledge of NOMA signals at GSSK users, and present a novel, energy-based ML detector  for GSSK signals. Additionally, we derive  expressions for the pairwise error probability (PEP) and  bit error rate (BER) union bound of the proposed detector.  To summarize, the main contributions of this paper are as follows.
\begin{itemize}
	\item We propose a novel energy-based ML detection strategy for  the detection of GSSK signals, which does not require the knowledge of NOMA signals.  We further demonstrate  that the performance of N-GSSK users with energy-based ML detection asymptotically  converges to that of iN-GSSK.
	\item We derive a tight approximation for the pairwise error probability (PEP) of GSSK users over Rayleigh fading channels and establish the tightness of the approximation through Monte Carlo simulations and numerical results.
	\item We derive a novel expression for the overall BER of NOMA users, taking in account the   BER of the active antenna index detection.
	\item We evaluate the spectral efficiency of the proposed N-GSSK with imperfect SIC, and demonstrate that it  outperforms  the c-GSSK scheme.
	
\end{itemize}

\subsection{Organization}
The remainder of this paper is organized as follows. Section \ref{SecSysModel} presents the N-GSSK system model . The detection techniques of GSSK and NOMA users are investigated in Section \ref{SecDetn}. The performance analysis in terms of PEP, BER and spectral efficiency for GSSK and NOMA users is presented in Section \ref{PerAnalsis}. Section \ref{section-Result} validates the theoretical analysis through numerical  and Monte Carlo simulations results.  C oncluding remarks are provided in Section \ref{conclusion}.

\section{System Model} \label{SecSysModel}
We consider a network with a base station (BS) equipped with $\Nt$ transmit antennas, serving a group of single antenna-equipped users, denoted by $\uezero, \ueone, \uetwo, \ldots, \ueg$. As illustrated in Fig.~\ref{NOMATX}, the BS generates  sequences of independent bits, which are then mapped to  constellation points in a GSSK constellation diagram. In GSSK modulation, $\nt$  out of  $\Nt$ antennas made active at a given transmission slot\footnote{Note that when $n_t = 1$, GSSK reduces to SSK. Therefore, we refer to N-GSSK with $n_t = 1$ as N-SSK in the remainder of the paper.}. Without loss of generality, we assume that $\uezero$ represents the  GSSK user, whereas other users are assigned to  the NOMA network. Note that the non-zero elements in  the GSSK symbols represent the superposition coded symbols of NOMA users. 

At the receiving end, $\uezero$  employs  maximum-likelihood (ML) detection to find the indices of active antennas, and decode the corresponding transmitted symbols, see Fig.~\ref{NOMARX}. Since the information is encoded onto the  spatial-constellation diagram,  the decoder of  $\uezero$  searches over all possible active  antenna combinations to obtain an estimates of the antenna indices \cite{Ntontin_IEEEVT_2012}. The rest of the users, i.e., $\uegone, \uegtwo, \ldots, \uegm$, which are multiplexed in the power domain based on the principle of SC and fractional transmit power allocation (FTPA),  employ GSSK detection followed by SIC.  

In summary, in the N-GSSK scheme, the information corresponding to the GSSK user is spatially modulated over $\nt$ antennas, each of which carry a NOMA signal, which is intended to NOMA users. Hence, by accommodating  NOMA users in a GSSK system, the overall spectral efficiency of the considered network is improved.

Let the effective channel gain over $\nt$ active transmit antennas at user $\uezero$ be denoted by $\hzeroeff$. Without loss of generality, let the effective channel gains of NOMA users be sorted in an increasing order, and given as $\goneeff <  \gtwoeff < \ldots < \gMeff$\cite{Lina_IEEETC_2018}. The signal transmitted from the BS for $M$ NOMA users is given by
\begin{equation}
X=\sum_{m=1}^{M}\sqrt{\alpha_m \pnoma}~ z_m,
\end{equation}
where $\pnoma$ is the transmit power,  $\alpha_m > 0$ is the power allocation coefficient, $z_m$ is the transmitted symbol corresponding to  the $m^{\text{th}}$ NOMA user, and $\sum_{m=1}^{M}\alpha_m =1$. Note that the power allocation coefficients are sorted in a descending order. i.e.,  $\alpha_1>\alpha_2>\cdots>\alpha_M$. We assume the symbol $z_m$ is drawn from a constant-modulus constellation, such as phase shift keying (PSK) \cite{Proakis05edition}. As mentioned earlier, the superimposed symbol $X$ is transmitted through a specific combination out of $M_H= {N_t \choose n_t}$ different combinations (i.e., possible constellation points), which are selected based on a predefined GSSK antenna mapping rule. 

In  c-GSSK, a random sequence of independent bits is fed into a GSSK mapper, where groups of bits are mapped to symbols in the spatial constellation diagram, which in turns activates a set of  transmit antennas \cite{Su_IEEE_2015}. Typically, each active  antenna  transmits a constant signal $1/\sqrt{\nt}$. However, in the considered N-GSSK system, the symbol $X/\sqrt{(\nt)}$ is transmitted over $\nt$ antennas with a total transmit power of $\pnoma$, similar to the generalized spatial modulation (GSM)\cite{Younis_IEEEconf_2010}. In other words, the transmitted signal on the $\Nt \times 1$ antenna array is given by
\begin{equation}
\left[\frac{X}{\sqrt{\nt}} ~0~ \cdots  ~0~ \frac{X}{\sqrt{\nt}}~ \cdots ~0 \right],
\end{equation}
where the non-zero elements represent the  active antennas $\nt$. Clearly, the signal received at the GSSK user is given by \cite{jaydeepan_IEEETWC_2009}
\begin{align}
y_0 = \sqrt{\rhodash} \ehjeff X + w_0,
\end{align}
where $w_0$ is assumed to be a circularly symmetric complex Gaussian (CSCG) random variable with zero mean and variance 
$\varn$ -- denoted by $\mathcal{CN}(0,\varn)$ and $\ehjeff=h_{j,1} +h_{j,2}+...+h_{j,n_t}$, where $j \in \{1,2, ...,M_H\}$  denotes the antenna combination for a given GSSK symbol.  The fading coefficients $h_{j,i}$ represent the channel gains from  the $i^{\rm th}$ active antenna  to the GSSK user, where each of which is modeled as complex Gaussian random variable with zeros mean and unit variance, i.e., $\mathcal{CN}(0,1)$.

Without loss of generality, we assume that $\varn = 1$. Also, $\rhodash \triangleq \rho/\nt$, where $\rho$ is the average signal-to-noise ratio (SNR). As explained earlier, the GSSK user  decodes its transmitted symbols by estimating the antenna indices using an ML decoder.  NOMA users, on the other hand, performs  GSSK decoding as wel as SIC to decode their own messages. In the considered system model, it is to be noted that although NOMA users receive  superimposed signals from multiple antennas, they do not exploit any of the known MIMO-NOMA techniques \cite{Zeng_IEEEWCL_2017}. 

\section{Detection Strategies} \label{SecDetn}
In this section, we propose an energy-based ML detector for  GSSK detection, which also used for  antenna indices estimation required by  NOMA users. 
\subsection{Energy-Based ML Detector for the GSSK User} \label{SubSecMLGSSK}
In iN-GSSK,   GSSK users are assumed to have  perfect knowledge of the NOMA signal $X$. Subsequently, conventional ML decoding is used to estimate the set of active antennas \cite{Kim_IEEE_CL_2018}. However, this assumption is unrealistic and impractical. Motivated by this, our proposed decoder first estimates   the energy of the received signal as follows:
\begin{align}
& \yE \triangleq \left | y_0 \right |^2=(\sqrt{\rhodash} \ehjeff X + w_0)(\sqrt{\rhodash} \ehjeff X + w_0)^*   \nonumber \\
&~~~~~~~~~~~~~= \rhodash P \hjeff + \nonumber \\
& ~~~~~~~~~~~~~~~~ \underbrace{\sqrt{\rhodash} X \ehjeff w_0^*+\sqrt{\rhodash}X^* \ehjeff^* w_0 + |w_0|^2}_{\text{Interference + Noise}~\triangleq~W_0}\\
&~~~~~~~~~~~~= \rhodash P \hjeff + W_0. \label{signalenergyeqn}
\end{align}
The exact distribution of $W_0$ is intractable and difficult to obtain.  However, $W_0$ in \eqref{signalenergyeqn} can be approximated as a complex Gaussian random variable with mean $\muI$ and variance $\sigmaI^2$, that is, $\mathcal{N}(\muI, \sigmaI^2)$. It can be readily shown that $\muI = \varn$, and $\sigmaI^2 = \varn[2 \rhodash \pnoma \nt +\varn]$. Therefore, the proposed energy-based ML detector at the GSSK user estimates the active antenna indices as follows:
\begin{align}
\hat{\mathbf{k}} = \underset{k \in \{1,2, ...,M_H\}}{\arg\min}\left \| \yE-\rhodash\pnoma \hkeff - \muI  \right \|^2,
\label{MLdetectorGSSK}
\end{align}
where $\hat{\mathbf{k}}$ represents the estimated antenna index vector. It can be readily noticed that the GSSK user does not  require the  knowledge of the superimposed NOMA signal. 

\subsection{Detection of NOMA Users}

As it can be inferred from Fig.~\ref{NOMARX},  GSSK symbols carry both NOMA signals and  active antenna indices. Therefore, it is crucial to correctly estimate the set of active antenna indices first in order to  reliably decode NOMA signals.  Towards this end,  NOMA users first perform energy-based  GSSK detection, followed by  conventional NOMA detection.

The key idea behind NOMA decoding is to employ SIC. In particular, the user with the highest allocated power decodes his own signal by treating  interference from other users' signals as noise. Other users
progressively cancel out the decoded signals of lower order users, i.e., users with higher power coefficients, and then decode their own signals while treating signals with lower power values as noise. 

The received signal at the $m^{th}$ NOMA user is given by \cite{Lina_IEEETC_2018}
\begin{align}
r_m=\egmeff X + n_m, ~~ m = 1, 2, \ldots, M
\end{align}
where, given the $j^{\rm th}$ set of antenna combination,  $\egmeff = g^m_{j,1}+...+g^m_{j,n_t}$ denotes the effective channel gain between the BS and the $m^{th}$ user, which is modeled as  zero mean complex Gaussian with unit variance and $n_m \sim \mathcal{CN}(0,1)$.  As noted earlier, the $m^{th}$ user performs SIC by decoding the signals of users with higher power, i.e., $\uegone, \uegtwo, \ldots, \uegmminusone$, 
while treating the signals  of   $\uegmplusone, \uegmplustwo, \cdots, \uegm$, as interference.

\section{Performance Analysis} \label{PerAnalsis}
In this section, we present a thorough performance analysis of the detection strategies used by the GSSK and NOMA users, focusing on the error rate analysis and spectral efficiency.
\subsection{Bound on BER of the GSSK User}
An upper bound on the average BER performance of the GSSK users can be obtained through union bound as \cite{Proakis05edition}
\begin{equation}
\pegssk \leq \frac{1}{\mh}\sum_{j=1}^{\mh}\sum_{k=1, k \neq j}^{\mh}M{(j,k) P(x_j\rightarrow x_k)}
\label{BER1}
\end{equation}
where $\mh$ is the set of all possible index sets of active antennas. Note that  $b_H = \lfloor \log_2 \mh \rfloor$ is the number of bits that can be conveyed by choosing a set of active antennas. Also, $M{(j,k)}$ denotes the number of bits in error between the signals $x_j$ and $x_k$, and $P(x_j \rightarrow x_k)$ denotes the pairwise error probability (PEP), which represents the probability of erroneously decoding $x_k$ when  $x_j$ was transmitted.

Following the detection strategy proposed in \eqref{MLdetectorGSSK}, the PEP of the GSSK user conditioned on channel vector  can be written as
\begin{align}
& P(x_j\rightarrow x_k|\ehjeff,  \ehkeff) \nonumber \\ 
&~~=Pr\left (\left | \yE-\rhodash\pnoma \zeta -\muI \right |^2 \leq \left | \yE-\rhodash\pnoma \xi - \muI \right |^2\right )  \label{PEP1} \\
&~~=Pr\left (\yE > \frac{\rhodash\pnoma}{2}  \left [ \xi+\zeta  \right ] + \muI \right )  \label{pepgsskeqlable3}
\end{align} 
where $\xi = \hjeff$ and $\zeta = \hkeff$. Recalling  that $\yE = \rhodash P \hjeff + W_j$, and $W_j \sim \mathcal{N}(\varn, \varn [2 \rhodash \pnoma \nt + \varn])$, \eqref{pepgsskeqlable3} can be written as 
\begin{align}
&  P(x_j\rightarrow x_k|\ehjeff, \ehkeff)  = Q\left (\frac{\frac{\rhodash \pnoma}{2}d(j,k) }{\sqrt{\varn[2\rhodash\pnoma  \nt +\varn]}}\right ) \label{condPEP}
\end{align}
\noindent 
where $d(j,k) \triangleq \left \lvert  \xi - \zeta\right \rvert$, and $Q(.)$ is the complementary CDF of a standard Gaussian random variable \cite{Proakis05edition}.  The unconditional PEP in \eqref{condPEP} can be realized by averaging the conditional PEP over the PDF of  $\xi$ and $\zeta$. 
Noting that $\xi$ and $\xi$ follow the exponential distribution with parameter $\lambda = 1/\nt$, denoted by $\texttt{Exp}(\lambda)$,   the exact PEP of GSSK users can be written as

\begin{align}
 P(x_j\rightarrow x_k)=\int_{0}^{\infty \nonumber }\int_{0}^{\infty}P(x_j\rightarrow x_k|\ehjeff, \ehkeff) ~  \nonumber \\
~~~~~~~~~~~~~~\times f(\xi)f(\zeta) ~d\zeta~ d\xi,   \label{doubleint11}
\end{align}
where $f(\cdot)$ denotes the exponential PDF with parameter $\lambda$. The PEP expression can be further rewritten as
\begin{align}
& P(x_j\rightarrow x_k)=\int_{0}^{\infty \nonumber }\int_{0}^{\infty} Q\left (  \frac{\left | \frac{\rhodash P \xi}{2}-\frac{\rhodash P \zeta}{2} \right |}{\sqrt{\varn(2 \rhodash \pnoma \nt +\varn)}}\right ) \nonumber \\
& ~~~~~~~~~~~~~~~~~~~~~~~~~~~~~~ \times \left(\frac{1}{\nt}\right)^2 \exp^{-\frac{(\zeta + \xi)}{\nt}} d \xi d\zeta. \label{doubleintgapproxpep}
\end{align}
The solution to  \eqref{doubleintgapproxpep} is presented in the following proposition.

\begin{Proposition}\label{Prop1}
The PEP of GSSK users is given by \eqref{PEP3} on the top of the next page, where $a \triangleq \dfrac{\rho'P\nt}{2\sigma_{n}\sqrt{2\rho'P\nt+\sigma_{n}^{2}}}$, and $G_{m,n}^{p,q}\left(\cdot \left \lvert \begin{array}{c}
- \\
- \end{array} \right. \right)$ represents the Meijer G-function.
\end{Proposition}   
\begin{figure*}[ht]
\begin{align}
& P(x_j\rightarrow x_k)=\dfrac{1}{2}\left[1-\frac{a}{\pi}\left\{ \begin{array}{c}
G_{2,2}^{1,2}\left(2 a^{2}\left \lvert \begin{array}{c}
\frac{1}{2},0;-\\
0;-\frac{1}{2}
\end{array} \right.\right)+\dfrac{1}{4}\sum_{k=0}^{\infty}\dfrac{1}{k!}G_{3,2}^{1,3}\left(\frac{a^{2}}{2} \left \lvert \begin{array}{c}
\frac{1}{2},-\frac{k}{2},-\frac{1}{2}-\frac{k}{2};-\\
0;-\frac{1}{2}
\end{array}\right.\right)\end{array}\right\} \right]	\label{PEP3}
\end{align}
\hrulefill
\end{figure*}   

\begin{proof}
See Appendix.
\end{proof}

It is worth noting that the infinite series  in \eqref{PEP3} converges very fast, and about $20$ terms suffice to obtain an accuracy up to four decimal places. More details are discussed in Sec.~\ref{section-Result}.  The PEP  expression in \eqref{PEP3} constitutes the basic building block for the derivation of the union bound   given by  \eqref{BER1}.

\subsection{Bound on BER of NOMA Users}
As stated earlier,  NOMA users first perform energy-based GSSK detection, followed by conventional NOMA detection. Without loss of generality, we consider the first user as the farthest user. Therefore, the received signal at the first user, conditioned on the $j^{\text{th}}$ set of antenna combinations can be represented as 
\begin{equation}
y_1 = \egoeff \left(\sqrt{\alpha_{1}\pnoma}z_1+\sum_{m=2}^{M}\sqrt{\alpha_{m}\pnoma}z_m\right) + w_1
\end{equation}
where $\sum_{m=2}^{M}\sqrt{\alpha_{m}\pnoma} z_m$ represents the sum of interference terms corresponding to the signals for the other users. Therefore, the conditional PEP for the first user can be represented as \cite{Lina_IEEETC_2018}, 
\begin{align}
&  P(z_1 \rightarrow \hat{z}_1|\gone) = \nonumber \\
& \hspace{-0.2cm}   Q\left(\frac{\sqrt{\alpha_{1}\pnoma}\gone\left|\deltaone\right|^2+2\gone \text{Re}\left \{\deltaone\sum_{j=2}^{M}\sqrt{\alpha_j\pnoma}z_j^*\right\}}{\sqrt{2}|\deltaone|\varnnoma}\right),
\end{align}
where $\deltaone=(z_1-\hat{z}_1)$ and $\gone=\left \lvert \egoeff \right \rvert$.\footnote{We omit the notation $j$ for simplicity.} Note that the statistics of $\gone$ can be obtained from the order statistics of the Rayleigh distribution. The average PEP over the PDF of $\gone$ is given by \cite{Lina_IEEETC_2018}
\begin{align}
& P(z_1 \rightarrow \hat{z}_1)=\frac{1}{2}\left(1-\frac{\nu \sigmah}{\sqrt{2\beta^2+\nu^2\sigmah^2}}\right)
\end{align}
where
\begin{align}
\nu=\sqrt{\alpha_{1}\pnoma}\gone\left|\deltaone\right|^2+2\gone\text{Re}\left \{\deltaone\sum_{j=2}^{M}\sqrt{\alpha_j\pnoma}z_j^*\right \},
\end{align}
$\beta= \sqrt{2}\left|\deltaone\right|\sigma_n$, and $\sigmah^2= \mathbb{E}\gone^2$. Similarly, the conditional PEP of the $m^{\text{th}}$ user can be evaluated as in \cite[Eq.~(21)]{Lina_ICC_2018} as
\begin{equation}
P(z_m \rightarrow \hat{z}_m | \gm)=Q\left(\frac{\gm\etam}{\sqrt{2}|\deltam|\varnnoma}\right),
\label{PEPNOMAuser1}
\end{equation}
where $\gm=\left \lvert \egmeff \right \rvert$ and
\begin{align}
& \etam=\sqrt{\alpha_{m}\pnoma}\left|\deltam\right|^2\nonumber+2\text{Re}\left \{\deltam\sum_{j=m+1}^{M}\sqrt{\alpha_j\pnoma}\hat{z}_m^*\right\}\\ \nonumber 
&~~~~~~~~~~~~~~~~~~~~~~~~~ +\text{Re}\left \{\deltam\sum_{k=1}^{m-1}\sqrt{\alpha_k\pnoma}\delta_k^*\right\},
\end{align}   
where $\deltam=(z_m-\hat{z}_m)$ and $\deltak=(z_k-\hat{z}_k)$. Following an approach similar to above, the average PEP over the PDF of $\gm$ can be written as  
\begin{align}
&P(z_m \rightarrow \hat{z}_m)=\frac{M!}{\sigmah^2(m-1)!(M-m)!}\\ \nonumber
&~~~~~~~~~~\sum_{j=0}^{m-1} \binom{m-1}{j}\frac{(-1)^{2(m-1)-j}}{[M-m+j+1]}\\ \nonumber
&~~~~~~~~~~\left ( 1-\frac{\etam \sigmah}{\sqrt{\etam^2 \sigmah^2+[M-m+j+1](\sqrt{2}|\deltam|\varnnoma) ^2 } } \right ). \nonumber
\label{PEPNOMA}
\end{align}

Next, the  above PEP expression is used to calculate an upper bound on the BER as
\begin{align}
& \penoma \leq \frac{1}{B}\sum_{\ell=1}^{L}P(z_\ell^{(m)}) \nonumber \\ 
& ~~~~~~~ \times \hspace{-0.3cm} \sum_{\ell=1,~z_\ell^{(m)}\neq \hat{z}_\ell^{(m)}}^{L} \hspace{-0.3cm} q(z_\ell^{(m)} \rightarrow \hat{z}_\ell^{(m)}) P(z_\ell^{(m)} \rightarrow \hat{z}_\ell^{(m)}),
\end{align}
where $z_\ell^{(m)}$ and $\hat{z}_\ell^{(m)}$ are the symbols for the $m^{\text{th}}$ user, $L$ is the number of symbols, $B$ is the number of information bits in symbol $z_\ell^{(m)}$, $P(z_\ell^{(m)})$ is the probability of occurrence of $z_\ell^{(m)}$, and $q(z_\ell^{(m)} \rightarrow \hat{z}_\ell^{(m)})$ is the number of bit errors when $z_\ell^{(m)}$ is transmitted and $\hat{z}_\ell^{(m)}$ is detected. The overall BER across $M$ NOMA users depends on the BER  of the active antenna indices and NOMA detectors, and is given by
\begin{align}
\petotalnoma \leq (1 \hspace{-0.1cm} - \hspace{-0.1cm} \pegssk) \sum_{m=1}^M \penoma + \pegssk,
\end{align}
and the overall BER for the entire N-GSSK system is given by
\begin{align}
\petotal \leq (1 \hspace{-0.1cm} - \hspace{-0.1cm} \pegssk) \sum_{m=1}^M \penoma + 2 \pegssk.
\end{align}

\subsection{Sum Rate of N-GSSK}
The sum rate of the GSSK user is given by 
\begin{align}
\rategssk= (1-\pegssk) \left \lfloor \log_2 {\Nt \choose \nt} \right \rfloor. \label{eqrategssk}
\end{align}
For NOMA users, the $m^{\text{th}}$ user will detect the $n^{\text{th}}$ users' message such that $n<m$, by performing SIC. The message from the users $n>m$ will be treated as noise at the $m^{\text{th}}$ user. Accordingly, the average data rate at the $m^{\text{th}}$ user, $m=1,\ldots,M-1$, is given by 
\begin{align}
&  R_m= \int_0^\infty \log\left(1+\frac{ \gm^2 \alpha_m \pnoma}{ \gm^2 \sum_{i=m+1}^{M}\alpha_i \pnoma + \varnnoma^2 }\right) {f}(\gm)~ d\gm,
\end{align}
where ${f}(\gm)$ denotes the PDF of $\gm$. Similarly, the average rate at the $M^{th}$ user is given by
\begin{align}
&  R_{M}= \int_0^\infty \log\left(1+\frac{ \gM^2 \alpha_M \pnoma}{ \varnnoma^2 }\right)  {f}(\gM)~ d\gM,
\end{align}
where $\gM=\left \lvert \egMeff \right \rvert$, and ${f}(\gM)$ denotes the PDF of $\gM$. Therefore, the average sum rate achieved by all $M$ NOMA users is given by
\begin{align}
& \ratenoma = \sum_{m=1}^M (1-\penoma) R_m,
\end{align}  
and the total sum rate of N-GSSK system is given by
\begin{align}
& \ratengssk = \rategssk + \ratenoma,
\end{align}
which is an improvement over the rate achieved by the c-GSSK system, given in \eqref{eqrategssk}. It is worth noting that the computational complexity of the proposed N-GSSK is nearly the same as iN-GSSK, which is discussed in \cite{Kim_IEEE_CL_2018}.
\section{Simulations and Numerical Results} \label{section-Result}
\begin{figure}
	\centering
	\includegraphics[width=3.7in, height=3in]{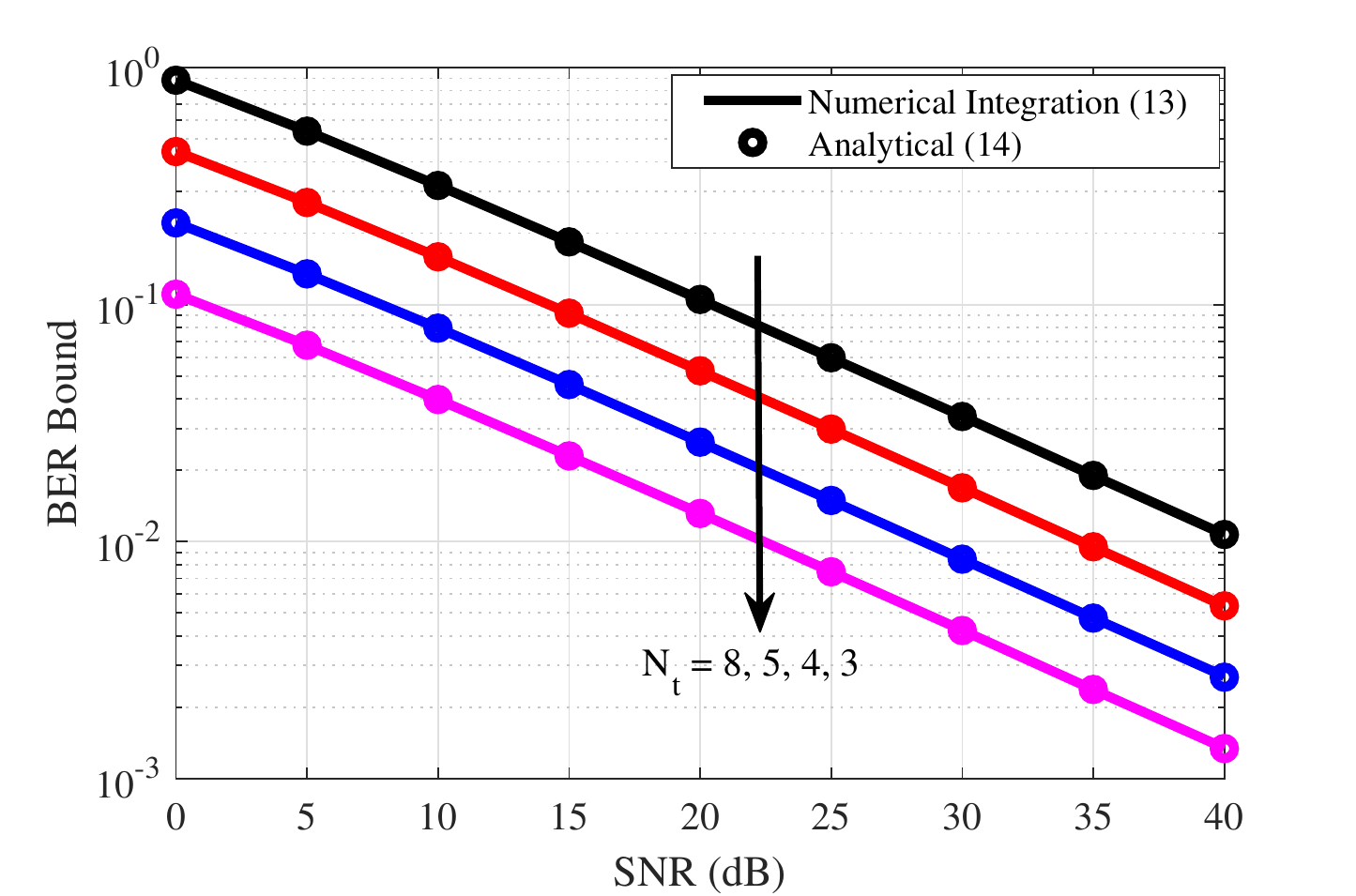} 
	\caption{Comparion between the union bound on BER calculated from \eqref{BER1}, obtained from simulated PEP from \eqref{doubleintgapproxpep}, and closed-form PEP \eqref{PEP3}, for different $N_t=8, 5, 4, 3$, with $\nt = 2$.} \label{BERGSSK}
\end{figure}

In this section, we analyze the performance of the proposed N-GSSK system in terms of BER and  spectral efficiency. Without loss of generality, we assume  a single GSSK user, i.e., $G=1$ and $M=2,3$ NOMA users. Fig.~\ref{BERGSSK} depicts the union bound performance given by \eqref{BER1} for $\Nt= 8$, $5$, $4$, and $3$, and $\nt=2$. The BER bound performance is evaluated using numerical integration, i.e., by numerically integrating \eqref{doubleintgapproxpep},  and based on the PEP expression derived in \eqref{PEP3}. 
It can be easily seen that there is a perfect match between the results of numerical integration and  the
results based on \eqref{PEP3}, validating the accuracy of the derived expressions.  It should be further emphasized that the infinite series in \eqref{PEP3} converges rather fast, where it has been noticed that truncating the series to the first twenty terms yields an accuracy up to four decimal places.

\begin{figure}
	\centering
	\includegraphics[width=3.7in, height=3in]{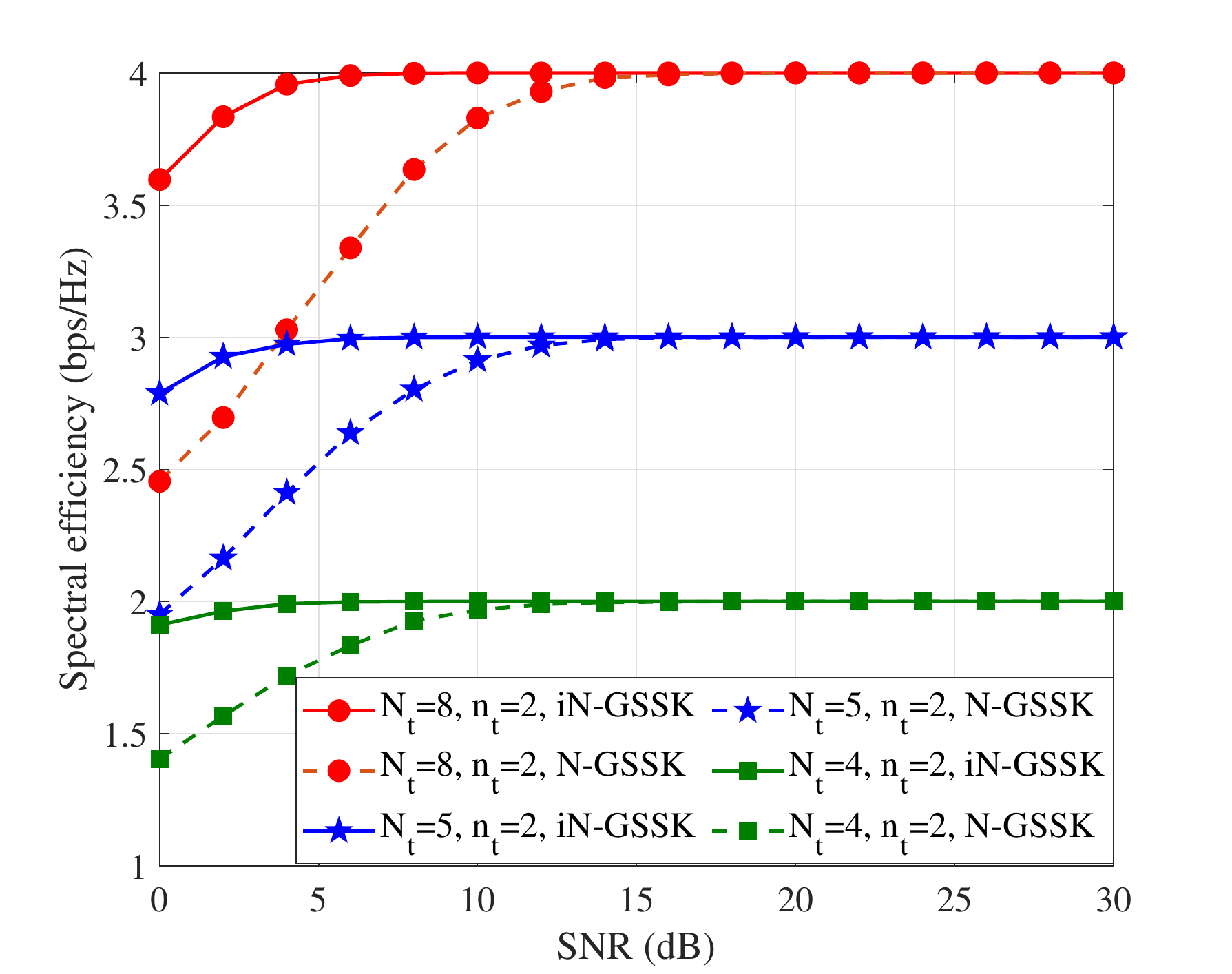} 
	\caption{Comparison of spectral efficiencies of N-GSSK and iN-GSSK.} \label{SEVsSNR}
\end{figure}
The spectral efficiency for the proposed N-GSSK, is compared with that of iN-GSSK in Fig.~\ref{SEVsSNR}. It is recalled here that in an iN-GSSK system, GSSK users are assumed to have the perfect knowledge of NOMA signals, which is  unrealistic, particularly in practical scenarios. Although, at low SNR values,  the spectral efficiency of the proposed N-GSSK is lower than that of iN-GSSK, it can be observed that both schemes  demonstrate similar performance at moderate-to-high SNR. Note that the performance loss of the N-GSSK is due to the suboptimality of energy-based ML detector. Furthermore, as shown, the spectral efficiency of the N-GSSK increases with increasing  $\Nt$ for a given $\nt$.

\begin{figure}
	\centering
	\includegraphics[width=3.7in, height=3in]{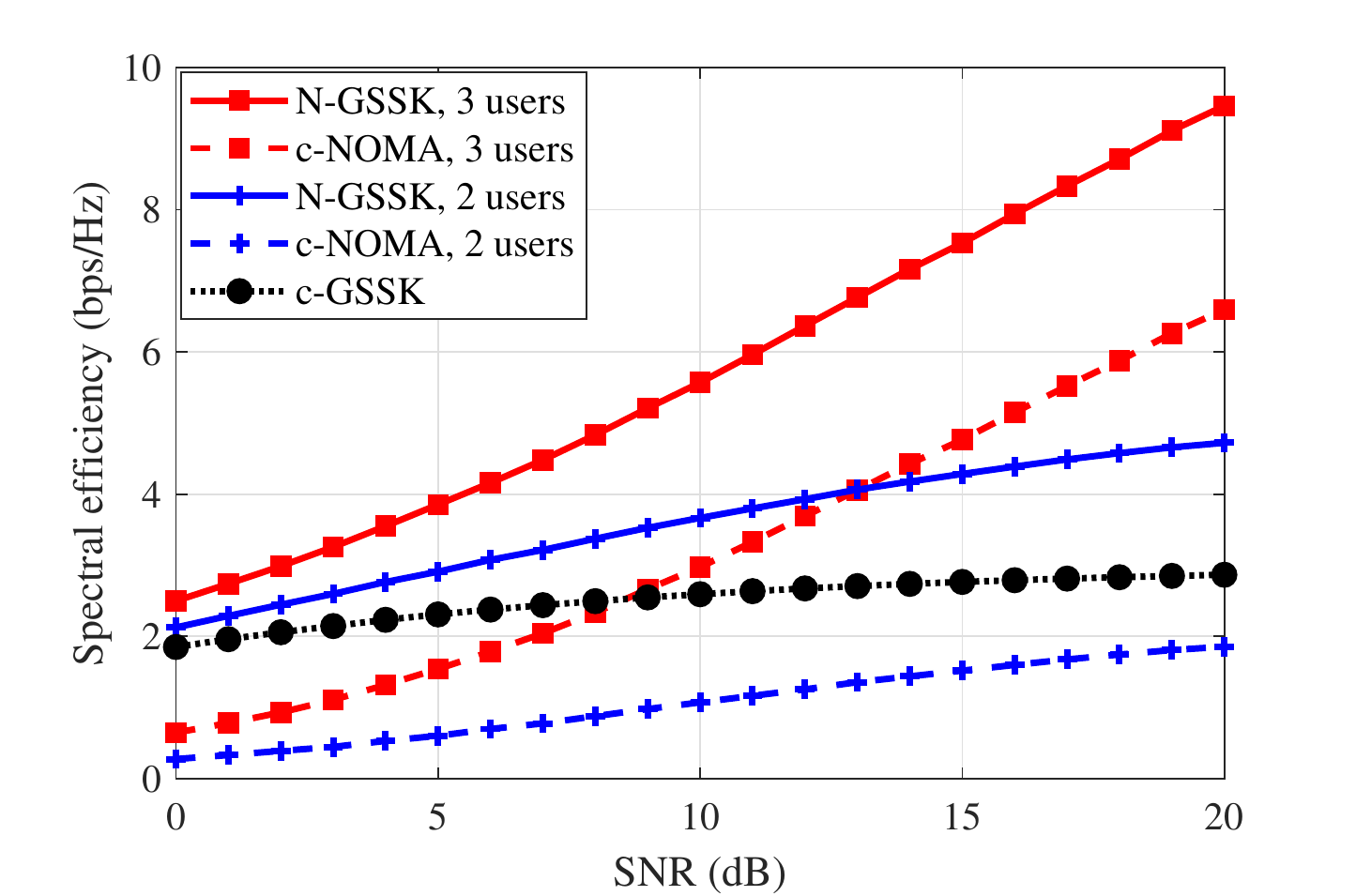} 
	\caption{Comparison of spectral efficiencies of N-GSSK, c-GSSK and c-NOMA for two and three users scenarios, with $\Nt=5$ and $\nt=3$. } \label{SEVsSNRonelambda}
\end{figure}

The spectral efficiency of N-GSSK, c-GSSK, and c-NOMA with $N_t=5$ and $n_t=3$ is presented in Fig.~\ref{SEVsSNRonelambda}. The power allocation coefficients are chosen as follows. For the two user scenario, $\alpha_1=0.8$, and $\alpha_2=0.2$, while for the three user scenario, $\alpha_1=0.7$, $\alpha_2=0.2$, and $\alpha_3=0.1$. It can be observed that N-GSSK  yields a better spectral efficiency  compared to c-GSSK, which further increases as the number of NOMA users increases. Additionally, the proposed N-GSSK yields a better spectral efficiency compared to c-NOMA, as it exploits the capacity gains due to both spatial and power domain multiplexing. 

\begin{figure}
	\centering
	\includegraphics[width=3.7in, height=3in]{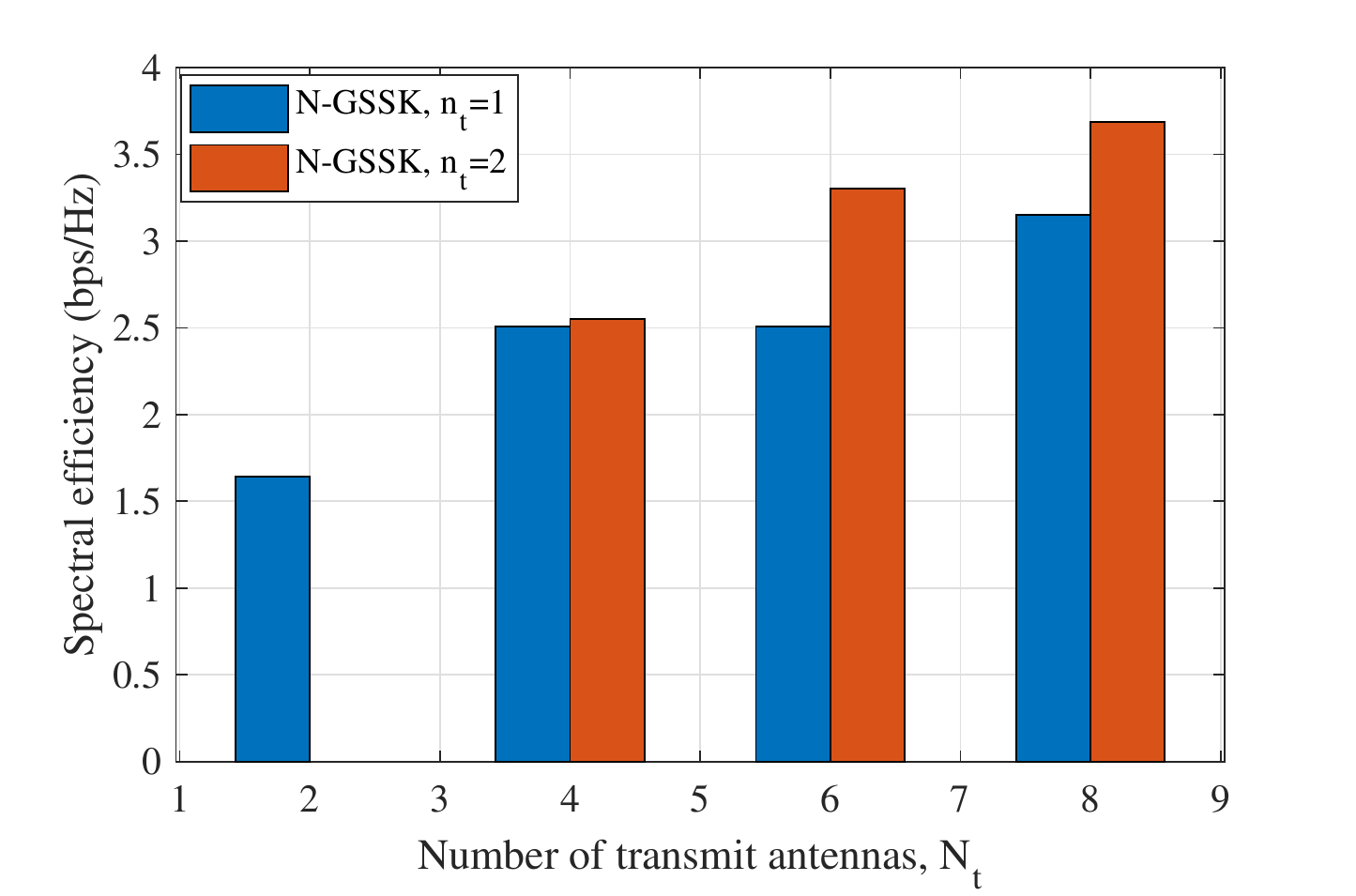} 
	\caption{Comparison of spectral efficiencies of N-GSSK (for $\nt$=2) and N-SSK (for  $\nt=1$) for different $\Nt$.} \label{barplot}
\end{figure}

Finally, Fig.~\ref{barplot} demonstrate the impact of the number of transmit antennas on the spectral
efficiency of N-GSSK. As mentioned earlier, N-SSK is a special case of N-GSSK, where $\nt=1$. For illustration purposes, we choose $\nt=2$ for N-GSSK. When $\Nt=2$, the performances of N-GSSK and N-SSK are equal, since the number of possible antenna indices are equal in both cases. As expected, N-GSSK outperforms N-SSK as $\Nt$ increases.

\section{Conclusion} \label{conclusion}
In this paper, we analyzed the spectral efficiency  of N-GSSK and proposed a novel, energy-based maximum likelihood detection scheme for N-GSSK. Unlike \cite{Kim_IEEE_CL_2018}, GSSK users in our setup can estimate  active antenna indices without the knowledge of NOMA signals. By  transmitting the superimposed NOMA signals and active antenna indices, we demonstrate that the spectral efficiency can be further enhanced. Furthermore, we investigated the performance of the proposed energy-based ML detector  through the derivation of  closed-form expressions for the pairwise error probability and BER union bound.  Monte Carlo simulations and numerical results were presented in order to  corroborate the analysis and establish the accuracy of derived expressions. Finally, we demonstrated that the N-GSSK scheme achieves significant spectral efficiency improvement, as opposed to the c-GSSK.

\section{Appendix: Proof of Proposition  \ref{Prop1}}
Let us define $x \triangleq \frac{\xi}{\nt}$, and $y \triangleq \frac{\zeta}{\nt}$ in \eqref{doubleintgapproxpep}. Therefore
\begin{align}
P(x_j\rightarrow x_k) = \int_{0}^{\infty}\int_{0}^{\infty}Q\left(a \left\vert x-y\right\vert \right)e^{-x}e^{-y}dxdy.
\end{align}
Noting that
\begin{align}
Q(| z |) = \frac{1}{2} \left[1-\erf\left(\frac{|z| }{\sqrt{2}}\right)\right],
\end{align}
the above integral can be written as
\begin{align}
& P(x_j\rightarrow x_k)=\frac{1}{2}\left[\underset{\triangleq \mathcal{I}_{1}}{\underbrace{\int_{0}^{\infty}e^{-x}dx\int_{0}^{\infty}e^{-y}dy}} \right. \nonumber\\
& ~~~~~~~~~~~~~~ \left. -\underset{\triangleq \mathcal{I}_{2}}{\underbrace{\int_{0}^{\infty}e^{-x}dx\int_{0}^{\infty}\erf\left(\gamma\left\vert x-y\right\vert \right)e^{-y}dy}}\right], \label{Pg1}
\end{align}
where $\gamma = a/\sqrt{2}$. It is easy to see that $\mathcal{I}_{1} = 1$. Also,
\begin{align}
& \mathcal{I}_{2}=\frac{1}{\sqrt{\pi}2\pi j}\oint_{\mathcal{C_{\textrm{s}}}}\frac{\Gamma(s)\Gamma\left(\frac{1}{2}-s\right)}{\Gamma\left(\frac{3}{2}-s\right)}\gamma^{-2s+1}ds \nonumber \\
& ~~~~~~~~~~~~~~ \left[\int_{0}^{\infty}e^{-x}dx\int_{0}^{\infty}e^{-y}\left\vert x-y\right\vert ^{-2s+1}dy\right], \label{I2}
\end{align}
where $\mathcal{C_{\textrm{s}}}$ is an appropriately chosen complex contour, ensuring the convergence of the above Mellin-Barnes integral. Substituting $z = \frac{y}{x}$, and further simplification yields \eqref{IntEqn}.
\begin{figure*}[ht]
\begin{align}
& \mathcal{I}_{2} = \frac{1}{2 \sqrt{\pi^3}  j}\oint_{\mathcal{C_{\textrm{s}}}}\frac{\Gamma(s)\Gamma\left(\frac{1}{2}-s\right)}{\Gamma\left(\frac{3}{2}-s\right)}\gamma^{-2s+1}ds\int_{0}^{\infty}x^{-2s+2}e^{-x}dx\int_{0}^{\infty}e^{-zx}\left\vert 1-z\right\vert ^{-2s+1}dz\nonumber \\
& \phantom{\mathcal{I}_{2}} = \frac{1}{\sqrt{\pi}}\left[\begin{array}{c}
\underset{\triangleq \mathcal{I}_{2}^{(1)}}{\underbrace{\frac{1}{2\pi j}\oint_{\mathcal{C_{\textrm{s}}}}\frac{\Gamma(s)\Gamma\left(\frac{1}{2}-s\right)}{\Gamma\left(\frac{3}{2}-s\right)}\gamma^{-2s+1}ds\underset{\triangleq \mathcal{J}_{2}}{\underbrace{\int_{0}^{1}\left(1-z\right)^{-2s+1}\underset{\triangleq \mathcal{J}_{1}}{\underbrace{\int_{0}^{\infty}x^{-2s+2}e^{-x(z+1)}dx}dz}}}}} \\
+\underset{\triangleq \mathcal{I}_{2}^{(2)}}{\underbrace{\frac{1}{2\pi j}\oint_{\mathcal{C_{\textrm{s}}}}\frac{\Gamma(s)\Gamma\left(\frac{1}{2}-s\right)}{\Gamma\left(\frac{3}{2}-s\right)}\gamma^{-2s+1}ds\underset{\triangleq \mathcal{J}_{4}}{\underbrace{\int_{0}^{\infty}x^{-2s+2}e^{-x}dx\underset{\triangleq \mathcal{J}_{3}}{\underbrace{\int_{1}^{\infty}\left(z-1\right)^{-2s+1}e^{-xz}}dz}}}}}
\end{array}\right] \label{IntEqn}
\end{align}
\hrulefill
\end{figure*}
Using \cite[Eqs.~3.381.4, 3.194.1]{refGradBk} for simplification, we get
\begin{align}
& \mathcal{J}_{1} =\frac{\Gamma\left(-2s+3\right)}{(z+1)^{-2s+3}}, \label{IntEqn1} \\
& \mathcal{J}_{2} = \Gamma\left(-2s+3\right)\int_{0}^{1}(z+1)^{2s-3}\left(1-z\right)^{-2s+1}dz \nonumber \\
& \phantom{\mathcal{J}_{2}} = 2^{2s-3}\Gamma\left(-2s+3\right)\int_{0}^{1}u^{-2s+1}\left(1-\frac{u}{2}\right){}^{2s-3}du \nonumber \\
& \phantom{\mathcal{J}_{2}} = {\frac{2^{2s-3} \Gamma\left(-2s+3\right)}{-2s+2}} \nonumber \\
& ~~~~~~~~~~~ \times \text{ }_{2}F_{1}\left(-2s+3,-2s+2;-2s+3;\frac{1}{2}\right) \nonumber \\
& \phantom{\mathcal{J}_{2}} = 2^{2s-3}\Gamma\left(-2s+2\right)\text{ }_{1}F_{0}\left(-2s+2;\cdot;\frac{1}{2}\right), \label{IntEqn2}
\end{align}
where ${}_pF_q(\cdot)$ is the hypergeometric function \cite{refGradBk}. Substituting \eqref{IntEqn1} and \eqref{IntEqn2} into \eqref{IntEqn} and simplifying further gives
\begin{align}
& \mathcal{I}_{2}^{(1)} =\frac{1}{2\pi j}\oint_{\mathcal{C_{\textrm{s}}}}\frac{\Gamma(s)\Gamma\left(\frac{1}{2}-s\right)\Gamma\left(-2s+2\right)2^{2s-3}}{\Gamma\left(\frac{3}{2}-s\right)}  \nonumber \\
& ~~~~~~~~~~~~~~~~ \times \gamma^{-2s+1}\text{ }_{1}F_{0}\left(-2s+2;.;\frac{1}{2}\right)ds\nonumber \\
& \phantom{\mathcal{I}_{2}^{(1)}} = \frac{\gamma}{4\sqrt{\pi}}\sum_{k=0}^{\infty}\frac{1}{k!}\frac{1}{2\pi j} \times \nonumber \\
& ~~ \oint_{\mathcal{C_{\textrm{s}}}}\frac{\Gamma(s)\Gamma\left(\frac{1}{2}-s\right)\Gamma\left(1+\frac{k}{2}-s\right)\Gamma\left(\frac{3}{2}+\frac{k}{2}-s\right)}{\Gamma\left(\frac{3}{2}-s\right)}\gamma^{-2s}ds \nonumber \\
& \phantom{\mathcal{I}_{2}^{(1)}} = \frac{\gamma}{4\sqrt{\pi}}\sum_{k=0}^{\infty}\frac{1}{k!}G_{3,2}^{1,3}\left(\gamma^{2}\left\vert \begin{array}{c}
\frac{1}{2},-\frac{k}{2},-\frac{1}{2}-\frac{k}{2};-\\
0;-\frac{1}{2}
\end{array}\right.\right). \label{I21}
\end{align}
Similarly, using \cite[Eqs.~3.381.2]{refGradBk}, it can be shown that
\begin{align}
& \mathcal{J}_{3} = x^{2s-2}e^{-x}\Gamma\left(-2s+2\right), \label{IntEqn3} \\
& \mathcal{J}_{4} = \frac{\Gamma\left(-2s+2\right)}{2}. \label{IntEqn4}
\end{align}
Substituting \eqref{IntEqn3} and \eqref{IntEqn4} into \eqref{IntEqn} gives
\begin{align}
\mathcal{I}_{2}^{(2)} & =\frac{\gamma}{\sqrt{\pi}}\frac{1}{2\pi j}\oint_{\mathcal{C_{\textrm{s}}}}\Gamma(s)\Gamma\left(\frac{1}{2}-s\right)\Gamma\left(1-s\right)\left(2\gamma\right)^{-2s}ds\nonumber \\
 & =\frac{\gamma}{\sqrt{\pi}}G_{2,2}^{1,2}\left(\left(2\gamma\right)^{2}\left\vert \begin{array}{c}
\frac{1}{2},0;-\\
0;-\frac{1}{2}
\end{array}\right.\right). \label{I22}
\end{align}
Finally, substituting \eqref{I22} and \eqref{I21} into \eqref{IntEqn} and \eqref{Pg1}, and substituting for $a = \sqrt{2} \gamma$ yields \eqref{PEP3}.

\bibliographystyle{IEEEtran}
\bibliography{IEEEabrv,NOMA_GSSK_biblo}
\end{document}